\documentclass[aps,pra,superscriptaddress,twocolumn]{revtex4}
\usepackage{amsmath}
\usepackage{calligra}
\usepackage{amsfonts}
\usepackage{graphicx}
\usepackage{dcolumn}
\usepackage{stackrel,amssymb}
\usepackage{color}
\usepackage{ulem}

\usepackage{bm}

\usepackage{hyperref}
\hypersetup
{
colorlinks=true,
citecolor=blue
}

\begin{document}

\title{Coherent and radiative couplings through 2D structured environments}
\author{F. Galve}
\email{fernando@ifisc.uib-csic.es}
\affiliation{IFISC (UIB-CSIC), Instituto de F\'isica Interdisciplinar y Sistemas 
Complejos, Palma de Mallorca, Spain}
\author{R. Zambrini}
\email{roberta@ifisc.uib-csic.es}
\affiliation{IFISC (UIB-CSIC), Instituto de F\'isica Interdisciplinar y Sistemas 
Complejos, Palma de Mallorca, Spain}

\begin{abstract}
We study coherent and radiative interactions induced among two or more quantum units, by coupling them to two-dimensional lattices acting as structured environments.
This model can be representative of atoms trapped near photonic crystal slabs, trapped ions in Coulomb crystals or to surface acoustic waves on piezoelectric materials,
cold atoms on state-dependent optical lattices, or even circuit QED architectures, to name a few. We compare coherent and radiative contributions for the isotropic and 
directional regimes of emission into the lattice, for infinite and finite lattices, highlighting their differences and existing pitfalls, e.g. related to long-time or
large-lattice limits. We relate the phenomenon of directionality of emission with linear-shaped isofrequency manifolds in the dispersion relation, showing a simple way
to disrupt it. For finite lattices, we study further details as the scaling of resonant number of lattice modes for the isotropic and directional regimes, and relate
this behavior with known van Hove singularities in the infinite lattice limit. Further we export the understanding of emission dynamics with the decay of entanglement
for two quantum, atomic or bosonic, units coupled to the 2D lattice. We analyze in some detail completely subradiant configurations of more than two atoms, which can 
occur in the finite lattice scenario, in contrast with the infinite lattice case. Finally we demonstrate that induced coherent interactions for dark states are zero for 
the finite lattice.
\end{abstract}

\maketitle

Engineering the coherent interaction between many quantum units is a key ingredient to simulate quantum phases of many-body systems, whereas radiative interactions is
instrumental to producing dark states and generating nonclassical states of light.
One of the most advanced platforms for quantum simulators are e.g. 2D Coulomb crystals of trapped ions \cite{thompson,Bollinger1,Bollinger2}, with
vibrational quanta of the ions' motion being used as a channel to effectively induce spin-spin interactions with tunable 
distance dependence. Excellent control of interactions between atoms and light in 1D waveguides has been recently shown \cite{painter2014}, leading to observation
of many atoms superradiance \cite{painter2015}. This has spurred theoretical proposals to use the engineered properties of light to mediate
strong long-range atom-atom interactions both in 1D \cite{chang2015} and 2D photonic crystal (PC) lattices \cite{tudela2015,tudela2016, galve2018}. 
Surface acoustic waves (SAW) on piezoelectric materials \cite{giedke2015} have been recently proposed as mediators between e.g. quantum dots, trapped ions, nitrogen-vacancy centers, or superconducting qubits.
Control of mediators, be it phonons in the ion and SAW cases, or photons in the PC case, is thus of key importance to shape interactions
to build future quantum simulators, as recently proposed for many-body physics \cite{chang2015}. \\
\indent Structured environments, in addition to their usefulness to shape coherent interactions, are a good playground to understand effects of collective dissipation (radiative couplings)
of distant units depending on the properties of the substrate. Understanding decoherence and dissipation in realistic scenarios is of utter 
importance for the successful realization of quantum technologies. This is for example the case when miniaturizing ion traps \cite{wine1} below 
the micrometer level, where details of decoherence sources, such as possible correlation lengths among adatoms' dipoles \cite{dipoles,Galve2017} 
could become accessible \cite{wine2}. Architectures of two or more units dissipating collectively 
can lead to super/sub-radiance \cite{Dicke,Haroche} and allow certain degrees of freedom to become noiseless 
\cite{noiseless,pazroncaglia}, which is of interest to realizing noise-free operations. 
Radiative coupling between bosonic probes enabling collective dissipation have been reported 
in cubic or triangular lattices, either isotropically decaying, or purely directional and long-range 
\cite{Galve2016}. This
translates directly into sub/super-radiant dynamics, as compared to independent decay rates (no radiative interaction) between quantum units. 
Long range radiative couplings not only expand the toolbox of quantum optics, but have important practical implications for quantum networking \cite{qinternet1,qinternet2} or
quantum memories \cite{qmem1,qmem2,qmem3}. Furthermore, understanding decoherence and dissipation in a broader class of platforms is of utter 
 importance for the successful realization of quantum technologies, where environment microscopic details play a key role, like for instance for  miniaturized ion traps
 \cite{wine1,dipoles,Galve2017}.\\
\indent Periodic structures can be described as tight-binding models for photons or phonons, 
disregarding the specific details of the materials used, and so allowing to study fundamental features. 
We focus on 2D structures because they allow richer features than 1D lattices, and can be implemented in
e.g. PC lattices \cite{tudela2015}, 3D arrays of (evanescently-coupled) light waveguides whose propagation direction $z$ represents the 
flow of simulated time \cite{Sciarrino1,Sciarrino2}, 2D Coulomb crystals of trapped ions \cite{thompson,Bollinger1,Bollinger2} or superconducting circuits \cite{cQED0,cQED1,cQED2,cQED3} to name a few.\\
\indent 
In this work we characterize coherent and radiative couplings induced by a 2D periodic (squared lattice) structure, both infinite and finite, highlighting their similarities
and differences. The paper is organized as follows: in section {\bf I} we introduce the model and its description in terms of a time-dependent master equation for the infinite lattice case, where we show the 
behavior of coherent and radiative contributions for isotropic and directional regimes. We analyze finite-time coefficients, explaining their meaning, and also the relation of long-time
decay rates with known van Hove singularities \cite{Tudela1,Tudela2,vanHove53}. We analyze the phenomenon of directional emission, showing how it can be broken, and highlight some 
pitfalls related to the ordering of large-lattice and long-time limits. An analysis on the validity of the infinite lattice limit is given. The decay dynamics of
entanglement between atomic or bosonic units due to the 2D structure is then studied, showing that it maps with the behavior expected from the cross-talk prediction.
In section {\bf II} we introduce the finite lattice, and analyze how the divergence of decay rates due to van Hove singularities are related to the number of resonant 
modes in the lattice as a function of time. We study how to construct dark states of two, three and four atoms from the form of the dissipation matrix, and show
that the form of the Lamb-shift matrix leaves dark states invariant. We finish by comparing exact dynamics of two-atom entanglement in the finite-lattice case, with the 
evolution under the master equation with time-dependent coefficients.

\begin{figure}[t!]
\includegraphics[width=\columnwidth]{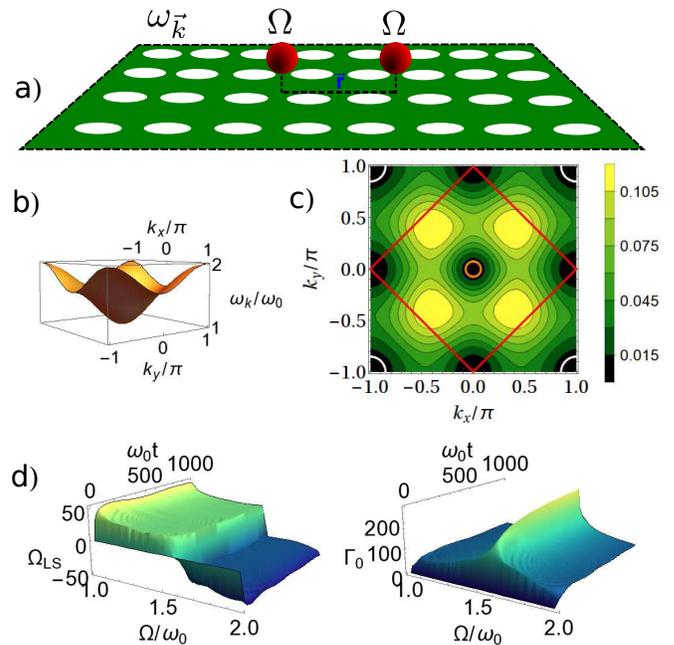}
\caption{a) Sketch of a periodic structured 2D lattice, e.g.
a PC. The system is either two harmonic or two-level probes with energy splitting $\Omega$
which couple locally to the structure at a relative distance $\vec{r}$. 
b) Lattice's (harmonic) dispersion relation $\omega_{\vec{k}}$, which regulates which manifold of excitations $\vec{k}_\Omega$ the two 
dissipative units couple to. c) Group velocity with $\omega_0=1$ and $J=3/16\omega_0^2$. We have drawn the iso-energy
manifold of resonant photon/phonon momenta $\vec{k}_\Omega$ for $\Omega=1.01\omega_0$ 
(orange), $\Omega=\sqrt{5/2}\omega_0$ (red) and $\Omega=1.98\omega_0$ (white).
d) Lamb-shift correction to the splitting $\Omega_{LS}$ and radiative 
decay rate $\Gamma_0$ induced by the lattice (without the factor $\frac{\lambda^2}{(2\pi)^2}$), with environment's frequencies in the range
$\omega_{\vec{k}}\in[1,2]\omega_0$.} 
\label{figDIBUJO}
\end{figure}

\section{Infinite lattice} We start by analyzing the interaction of a two-unit, bosonic or atomic, quantum system coupled to a structured
substrate such as that of Fig.~\ref{figDIBUJO} (there a PC lattice is represented). 
Our analysis holds for periodic, bosonic, linear  lattices (modeled by quadratic Hamiltonians).
For a square-lattice symmetry dispersion is either
\begin{equation}
\omega_{\vec{k}}=\omega_0-2J(\cos k_x +\cos k_y) 
\end{equation}
for coupled cavities or tight-binding models, or
\begin{equation}
\omega_{\vec{k}}=\sqrt{\omega_0^2+8J(\sin^2\frac{k_x}{2}+\sin^2\frac{k_y}{2})}
\end{equation}
for harmonic (spring-like) couplings. It can be seen that both dispersions have almost the same shape if they are scaled properly to fit in the range of values.
The tight-binding dispersion ranges from $\omega_0-4J$ to $\omega_0+4J$, whereas the harmonic one ranges from $\omega_0$ to 
$\sqrt{\omega_0^2+16J}$, so when we use the latter we set $J=3\omega_0^2/16$ so that $\omega_{\vec{k}}\in[1,2]\omega_0$ 
(as in Fig.~\ref{figDIBUJO}). Of course, we can offset all energies so that the frequency band lies e.g. $\omega_{\vec{k}}\in[-1,1]\omega_0$
for both dispersions, being only important the detuning between emitters and the resonant pseudomomentum manifold in the lattice.
The cases we study in this paper behave equally for both dispersion, because the implied resonant manifolds are the same, so we will
interchangeably use both, making the choice explicit only when we use the tight-binding one. \\
The two identical and independent emitters can be either bosons, $H_S=\Omega(a_1^\dagger a_1+a_2^\dagger a_2)$, or  two-level atomic systems (TLS), $H_S=\Omega(\sigma_1^+ \sigma_1^-+\sigma_2^+ \sigma_2^-)$.
We consider local coupling to the structure $$H_{\text{int}}=\lambda(a_1 A(\vec{r}_1)^\dagger+a_2 A(\vec{r}_2)^\dagger +h.c.)$$ or
$H_{\text{int}}=\lambda(\sigma_1^- A(\vec{r}_1)^\dagger+\sigma_2^- A(\vec{r}_2)^\dagger +h.c.)$ for TLS.
The  annihilation operators $A(\vec{r})$
for the light field (atoms interacting with PC) or surface displacement field (SAW case), are evaluated at the positions of the atoms (bosons) $\vec{r}_1$ and $\vec{r}_2)$.
Interactions extended in space over a range $\sigma$ (instead of local) are known to introduce frequency cut-off functions which limit interaction strength for lattice's frequencies
above $\sim 1/\sigma$ \cite{Galve2016}.
The interaction between system and lattice $\lambda$ is assumed to be weak, allowing for a
Born-Markov treatment \cite{breuer}. The dynamics, for environment at T=0, is given by
\begin{equation}
 \dot{\rho}_S=-i[\tilde{H}_S(t),\rho_S ]+\sum_{j,l=1}^2 \Gamma_{j l}(\vec{r},t) ( a_j \rho_S a_l^\dag - \frac{1}{2} \{a_l^\dag a_j,\rho_S\})\nonumber
\end{equation}
with $\tilde{H}_S(t)=H_S+H_{LS}(t)$ the Lamb-Shift corrected Hamiltonian, $H_{LS}(t)=\Omega_{LS}(t)(a_1^\dagger a_1+a_2^\dagger 
a_2)+\lambda_{LS}(\vec{r},t)(a_1a_2^\dagger+h.c)$ and $\vec{r}=|\vec{r}_1-\vec{r}_2|$. 
We define the cross-damping $\Gamma_c\hat{=}\Gamma_{12}=\Gamma_{21}$ and the 
self-damping $\Gamma_0\hat{=}\Gamma_{11}=\Gamma_{22}$ rates. Using $A(\vec{r})=(2\pi)^{-1}\int_{-\pi}^{\pi}d^2\vec{k}\ \exp(-i\vec{k}\cdot\vec{r})A(\vec{k})$  (i.e. infinite-lattice), the spatial dependent coefficients are
\begin{equation}\lambda_{LS}(\vec{r},t)=\frac{-\lambda^2}{(2\pi)^2}\int_{-\pi}^{\pi}d^2\vec{k}\ \frac{1-\cos[t(\Omega-\omega_{\vec{k}})]}{\Omega-\omega_{\vec{k}}}\cos(\vec{k}\vec{r})\end{equation}
\begin{equation}\Gamma_c(\vec{r},t)=\frac{2\lambda^2}{(2\pi)^2}\int_{-\pi}^{\pi}d^2\vec{k}\ \frac{\sin[t(\Omega-\omega_{\vec{k}})]}{\Omega-\omega_{\vec{k}}}\cos(\vec{k}\vec{r})\end{equation}
and the onsite coefficients are just $\Omega_{LS}(t)=\lambda_{LS}(0,t)$, $\Gamma_0(t)=\Gamma_c(0,t)$. They are the sum of counter-propagating plane waves, weighted by their resonance
with $\Omega$. The emission of only one unit is proportional to the same integral (note that due to the $\vec{k}\to-\vec{k}$ symmetry we can replace cos by an exponential), and thus the 
spatial features of $\Gamma_c$ give a precise idea of the radiation pattern of one emitter.
We retain, and analyze later, the time dependence of the coefficients in the master equation, which account for the build-up of a
communication channel between quantum units \cite{Benedetti}. This accounts to having performed Born and first-Markov approximations \cite{breuer}.
Usually in the literature the second-Markov approximation (long-time $t\to\infty$ limit) is used:
\begin{equation}
 \lambda_{LS}(\vec{r},\infty)=\frac{-\lambda^2}{(2\pi)^2}PV\int_{-\pi}^{\pi}d^2\vec{k}\ \frac{\cos(\vec{k}\vec{r})}{\Omega-\omega_{\vec{k}}}
\end{equation}

\begin{equation}
 \Gamma_c(\vec{r},\infty)=\frac{2\lambda^2}{(2\pi)^2}\int_{-\pi}^{\pi}d^2\vec{k}\ \delta(\Omega-\omega_{\vec{k}})\cos(\vec{k}\vec{r})
\end{equation}

The dynamics for identical uncoupled probes can be diagonalized at all times,
in the common and relative coordinates 
$a_\pm=(a_1\pm a_2)/\sqrt{2}$, resulting in two completely independent dynamics:
$$\dot{\rho}_S= \sum_{j=\pm}-i[H_j,\rho_S (t)]+\Gamma_{j}(\vec{r},t) ( a_j\rho_S a_j^\dag - \frac{1}{2} \{a_j^\dag a_j,\rho_S\})$$
with $H_\pm=(\Omega+\Omega_{LS}\pm\lambda_{LS})a_\pm^\dagger a_\pm$ and $\Gamma_\pm=\Gamma_0\pm\Gamma_c$.
The latter expression indicates that those modes can become noiseless when $\Gamma_c=\pm\Gamma_0$, which
is known to lead to preservation of entanglement at long times \cite{pazroncaglia,Galve2016}. When 
$\Gamma_c=+\Gamma_0$ we have $\Gamma_-=0$ and thus the mode $a_-$ is not dissipating, what is typically called a
common bath situation (CB). With $\Gamma_c=0$ both modes dissipate, the separate baths scenario (SB).
Otherwise, if $\Gamma_c=-\Gamma_0$ so we have $\Gamma_+=0$
and thus it is the center of mass motion which is noiseless \cite{Galve2017,zoller2008}. We termed this situation anti-common bath (aCB) \cite{Galve2017}.

The equivalent expressions for the two TLS case are obtained by substituting $a_j\to \sigma_j^-$ and $a_j^\dagger\to \sigma_j^+$.
The dynamics is then diagonal in the operator basis $\sigma_\pm^-=(\sigma_1^-\pm\sigma_2^-)/\sqrt{2}$, which has as 
fixed points $|00\rangle$ and $|\Psi_-\rangle=$ for CB ($\Gamma_c/\Gamma_0=1$) with definition $|\Psi_\pm\rangle=(|01\rangle\pm|10\rangle)/\sqrt{2}$. The state $|00\rangle$ is trivial in the sense that the lattice is a $T=0$
reservoir. However, if the system starts with $|\Psi_-\rangle$, entanglement is conserved asymptotically. For the aCB case ($\Gamma_c/\Gamma_0=-1$) it is
$|\Psi_+\rangle$ which is conserved, and none is conserved asymptotically when SB ($|\Gamma_c|<\Gamma_0$).

\subsection{Finite time coefficients} The coefficients $\Omega_{LS}$ and $\Gamma_0$, equal to their single-unit counterparts, can be seen in figure~\ref{figDIBUJO}. Both settle to constant
values rather fast, at $\omega_0 t\sim 10$, justifying the broadly used long-time limit master equations. 
The single-unit decay rate $\Gamma_0$ also settles quickly to a constant, except for the specific splitting $\Omega=\sqrt{5/2}\omega_0$ (from now on we will 
work with the harmonic dispersion, unless otherwise estated), where it has its highest value (for $J=3/16\omega_0^2$).
The decay rate at this particular $\Omega$ is divergent, see Fig.~\ref{figDIBUJO}d right, in the long time limit 
$\Gamma_0(t\to\infty)\propto\int_{-\pi}^{\pi}d^2\vec{k}\ \delta(\Omega-\omega_{\vec{k}})=\int_{\vec{k}_\Omega}d^2\vec{k}\ |\vec{\nabla}\omega(\vec{k}_\Omega)|^{-1}$.
It is the sum of all inverse group velocities (density of states, DoS, see fig. \ref{figDIBUJO}c) of the manifold of excitations $\vec{k}_\Omega$ resonant with
$\Omega$. For $\Omega=\sqrt{5/2}\omega_0$ (red line) it picks up points (black regions in plot) where the density of states diverges, known as van Hove singularities
\cite{vanHove53}. Due to the periodic structure of reciprocal space these singularities are known to appear {\it generically} \cite{Morse1925} and 
lead to logarithmic divergences in 2D. We also note that other points (not relevant for our work) have zero DoS, at the extrema of the band; here bound states form
and perturbative approaches are no longer valid \cite{wang90,quang94,intJ, boundCirac,boundCicca}. 
\begin{figure}
\includegraphics[width=\columnwidth]{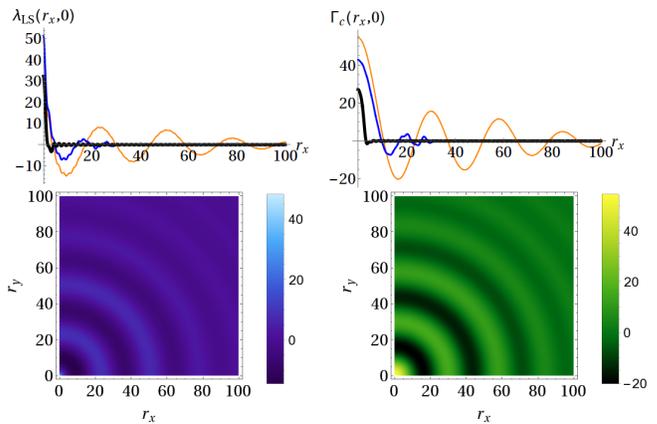}
\caption{2D lattice with $J/\omega_0=3/16$ so that $\omega_{\vec{k}}\in[\omega_0,2\omega_0]$,
isotropic case $\Omega=1.01\omega_0$. 
Top: Lamb Shift coupling (left) $\lambda_{LS}(\vec{r})$ and cross-damping (right) 
$\Gamma_c(\vec{r})$, normalized by the factor $\lambda^2/[(2\pi)^ 2]$, for different times 
$\omega_0 t=10, 100, 1000$ (black, blue, orange), along the line $r_y=0$. 
Bottom: Full space dependence of $\lambda_{LS}(\vec{r})$ (left) and $\Gamma_c(\vec{r})$ (right)
for $\omega_0 t= 1000$.
} 
\label{fig1}
\end{figure}
\begin{figure}
\includegraphics[width=\columnwidth]{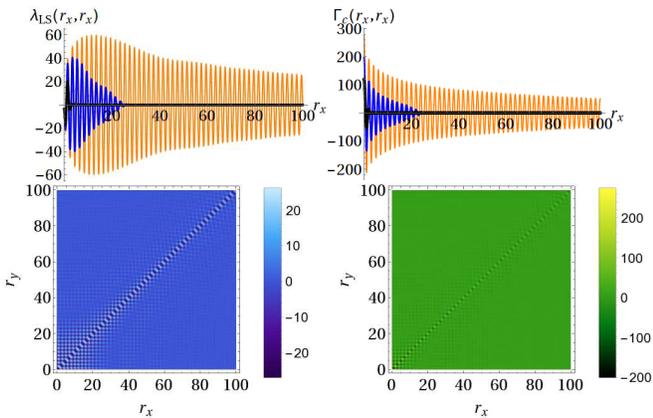}
\caption{Same quantities as in Figure~\ref{fig1}, but for the diagonal case 
$\Omega=\sqrt{5/2}\omega_0$. Figures at the top are drawn along the $r_x=r_y$ line.} 
\label{fig2}
\end{figure}
\begin{figure}
\includegraphics[width=\columnwidth]{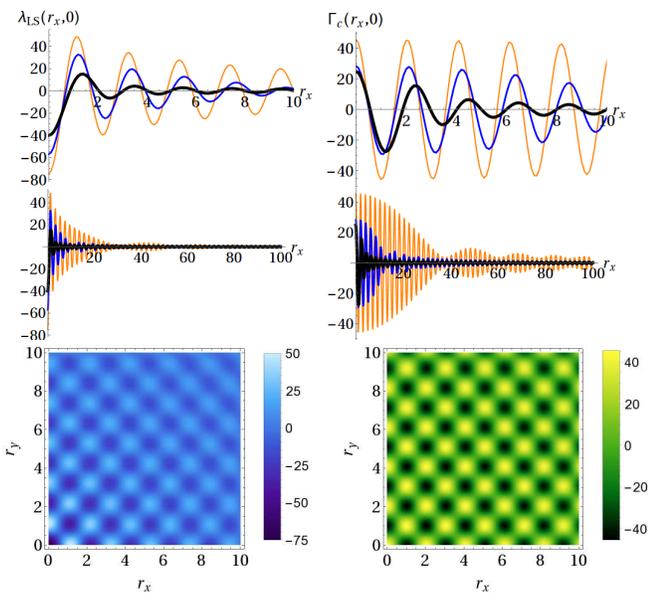}
\caption{Same quantities as in Figure~\ref{fig1}, but for the high-$k$
case $\Omega=1.999\omega_0$. Figures at the top (shorter times) and middle (longer times) are drawn along the $r_y=0$ line.} 
\label{fig3}
\end{figure}
\\
\indent The most important consequence of retaining time-dependent coefficients in the master equation, instead of just using their long-time limits (the usual Markovian
master equation), comes from the cross-talk. From figures \ref{fig1}, \ref{fig2} and \ref{fig3} we see that it takes a non-negligible time for the coherent and radiative interactions to become 
nonzero for distant units. During that time, the two quantum units see independent environments because $\Gamma_c=0$, and only when it becomes nonzero 
they start `seeing' a correlated environment. Take for example two units with $\Omega=1.01\omega_0$, as in figure~\ref{fig1}, at a distance $\sim 30$:
before $\omega_0 t=1000$ they have $\Gamma_c=0$ and thus decay independently, whereas from that time on they have $\Gamma_c\sim\Gamma_0/3$, leading to $\Gamma_-/\Gamma_+\sim1/2$.
This means that an initial state $|\psi_+\rangle$ will decay twice faster than  $|\psi_-\rangle$, but {\it only from that time onwards}.\\
\indent If the system frequency is increased reaching the middle of the band, fig. 3 ($\Omega=\sqrt{5/2}\omega_0$), it was predicted in \cite{2Dphotonic,Galve2016} that 
it leads to `diagonal-only' propagation of excitations, and thus to decay of cross-damping in all directions except for the diagonals. 
This behavior has been checked against the exact (atom-lattice case) dynamics (see \cite{galve2018,Tudela1,Tudela2}).\\
\indent We plot next the Lamb shift LS (coherent) coupling $\lambda_{LS}(\vec{r})$ and the cross-damping 
coefficient $\Gamma_c(\vec{r})$, without the irrelevant factor 
$\lambda^2/[(2\pi)^2]$ for the 3 cases highlighted in Fig.\ref{figDIBUJO}c,
corresponding to increasing frequency of the units interacting with infinite planar squared lattices. 
In the first case ($\Omega\simeq\omega_0$) the system is resonant 
with the low-momentum manifold of excitations in the lattice, 
corresponding to an approximately isotropic dispersion relation 
$\omega_{\vec{k}}\simeq \sqrt{\omega_0^2+2g|\vec{k}|^2}$, as can be seen in the 
spatial shape of the propagation in Fig. \ref{fig1}. 
The case of high{\it-k} ($\Omega=1.999\omega_0$)  in Fig. \ref{fig3} resonates with the `corners' of the dispersion relation (Fig.\ref{figDIBUJO}c) and, although apparently anisotropic, 
can be seen to lead also to isotropic decay of the master equation, with a superimposed alternating oscillation. 
Writing for one of the four corners $\vec{k}=(\pi-\delta_x,\pi-\delta_y)$, we have $\cos(\vec{k}\vec{r})=\cos[\pi(x+y)]\cos(x\delta_x+y\delta_y)$.
Integration of variables $\delta_{x,y}$ is again isotropic and thus leads to a $J_0(|\delta||\vec{r}|)$ decay, times the oscillating factor $\cos[\pi(x+y)]$.
Finally, the `diagonal' case ($\Omega=\sqrt{5/2}\omega_0$) is a 
peculiar case where the excitations conspire to produce a purely diagonal propagation \cite{galve2018} as displayed in  Fig. \ref{fig2}.\\
\indent At long times distant points are both coherently and radiatively coupled with different strengths
(orange curves respectively in Fig. \ref{fig1}, \ref{fig2},\ref{fig3}).
It is to be noted that the spatial dependence of 
coherent and incoherent contributions (for each case) are rather similar although not equal, see e.g.
the beating of long time Lamb-shift as compared to cross-damping (top fig.~\ref{fig2}), 
or their different decay behaviors (top fig.~\ref{fig3}), since both consist in integrating
$\cos(\vec{k}\vec{r})$ but with the different weights $[1-\cos(\Omega-\omega_{\vec{k}})]/(\Omega-\omega_{\vec{k}})$ 
and $\text{sinc}(\Omega-\omega_{\vec{k}})$. Both weights are most important when $\Omega\simeq\omega_{\vec{k}}$
but have different functional behavior, leading to these differences.

\subsection{Radiative directional coupling} 
\begin{figure}[h!]
\includegraphics[width=\columnwidth]{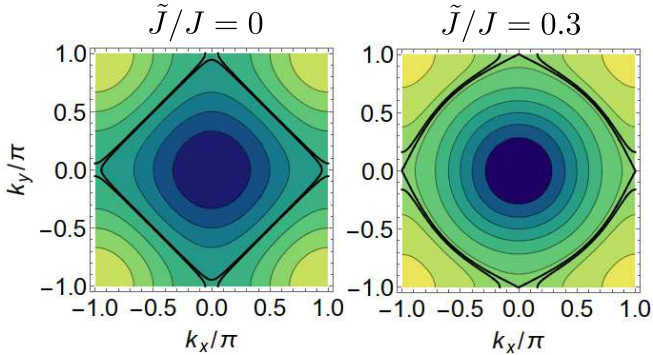}
\caption{Tight-binding dispersion relation without $\tilde{J}/J=0.0$ (left) and with $\tilde{J}/J=0.3$ (right) next-to-nearest neighbor hopping.
The manifold of interest, highlighted between darker contour lines, ceases to be straight and acquires a curvature.} 
\label{figDISPERSION}
\end{figure}

\begin{figure}[h!]
\includegraphics[width=0.8\columnwidth]{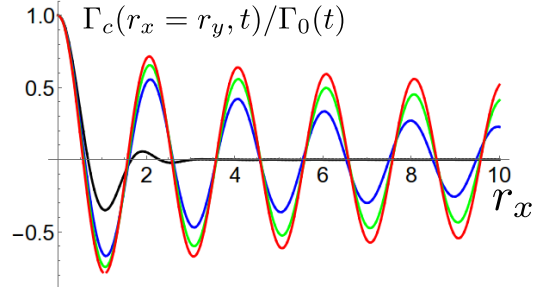}
\caption{Normalized cross-talk as a function of distance, with $J\cdot t/N=10$ (tight-binding dispersion), and $N=20,200,1000, 5000$ (black, blue, green, red) } 
\label{figtN}
\end{figure}

The `diagonal' ($\Omega=\sqrt{5/2}\omega_0$ or $\Omega=\omega_0$ for the tight-binding dispersion relation) 
was highlighted in \cite{Galve2016} as displaying purely directional radiative coupling (also radiation by one emitter).
This happens for cubic and triangular lattices \cite{Galve2016}, and also their higher-dimensional analogues, as well as for  graphene \cite{graphene}, 
being then present in common geometric configurations. It has been recently explained as a consequence of linear-shape iso-frequency manifolds
in the dispersion relation \cite{galve2018}: a rotation into that momentum coordinate selects a specific value of that momentum, which produces
a plane wave (or sine) in that direction, and is thus non-decaying. In the orthogonal coordinate we still sum many momenta, which gives generically
an incoherent sum of waves and thus a decaying function along the corresponding spatial coordinate. A simple mechanism breaking the linear shape 
for a square-symmetric lattice was proposed in \cite{galve2018}, consisting in adding a hopping term between diagonal neighbors in the lattice.
This results in e.g. a modified tight-binding dispersion relation $\omega_{\vec{k}}=\omega_0-2J(\cos k_x +\cos k_y)-4\tilde{J}\cos k_x \cos k_y$,
whose last term gives curvature to the isofrequency manifold. This is easily seen in Fig.~\ref{figDISPERSION}.
Because of the van Hove singularities in this manifold, the case of directional emission allows inspection of some possible pitfalls related
to the ordering of the large-lattice and long-time limits.
If we first obtain the analytic shape of $\Gamma_c(\vec{r})$ for periodic boundary conditions by taking a finite lattice 
with $t\to\infty$ (which exactly selects $N/2$ modes on each of the red straight lines in fig.~\ref{figDIBUJO}c) and taking then the 
large-lattice limit $N\to\infty$, which yields $\Gamma_c(\vec{r})/\Gamma_0=(x\sin\pi x-y\sin\pi y)/[\pi(x^2-y^2)]$, i.e. 
$[{\rm sinc} (\pi |\vec{r}| )+\cos(\pi |\vec{r}|)]/2$ at the diagonals and $\Gamma_c(\vec{r})/\Gamma_0=\text{sinc}(\pi r_y)$ at $r_x=0$.
This function has a maximum value $1$ at $|\vec{r}|=0$ and after the sinc decays, it has $1/2$ at the diagonals and $0$ elsewhere.
It is a peculiar fact that interchanging the limits gives a different result. By setting $t<N$ and doing $N\to\infty$ in fact gives 
$\Gamma_c(\vec{r})/\Gamma_0=1$. At intermediate $N$ we see a shape similar to directional radiation but with a region near $\vec{r}=0$ with almost
maximum cross-talk (see figure ~\ref{figtN}), i.e. higher than the $1/2$ value obtained when $t\gg N$. In addition, its absolute value is logarithmic-divergent,
as is well-known from van-Hove singularities \cite{vanHove53}. The very different behavior can be explained as follows: in the first case, $t\gg N$, only 
$N/2$ modes are resonant, because the spacing in momentum space goes as $\Delta k\sim 1/N$ and $t$ is so large that the $sinc$ function selects a momentum manifold always smaller
than $\Delta k$. In that way, we only sum the contribution of a finite number of modes. In spite of that, see later discussion on the scaling of the decay rate,
the cross-talk diverges as $N\ t$. In the second case, it is even worse: for $t\ll N$ the amount of resonant modes selected by the $sinc$ is even bigger, and does not have only
contributions from the momentum manifold yielding directionality; it has more contributions, and this is reflected in the modified cross-talk shape in Fig.~\ref{figtN}.

\subsection{Timescales and modes choice}
A comment is in order about timescales. The possibility to use exponential waves as ansatz for guided modes in a lattice totally disregards that lattices 
are always finite, an assumption that can be perfectly valid within a range of parameters. Take for example the tight-binding dispersion and note 
that the highest group velocity $d\omega_{\vec{k}}/d\vec{k}$ (see Fig.~\ref{figDIBUJO}c) is $2\sqrt{2}J$. Once any signal reaches the boundary of the
lattice, bouncings and revivals of the system dynamics are expected to occur. This happens for times $J\ t\sim N$. The decay dynamics of the system 
is however governed by the system-lattice coupling $\lambda$, yielding a decay rate $\Gamma_0\sim \lambda^2$, and thus significant decay is expected 
to happen at times $\sim 1/\lambda^2$. Thus the exponential wave ansatz is valid whenever the characteristic decay time is at least smaller than the revival 
of signals at the boundaries. This implies $1/\lambda^2< N/J$, and thus requires $\lambda^2>J/N$. Taking $J$ as energy scale,
and a lattice of $N\sim 100$ sites, already implies $\lambda\sim 0.1$, within the strong coupling regime, as in \cite{Tudela1}.
For this reason it is advisable to use finite-lattice mode functions, as we do in later sections. 

\subsection{Entanglement decay} 
\begin{figure}[h]
\includegraphics[width=0.7\columnwidth]{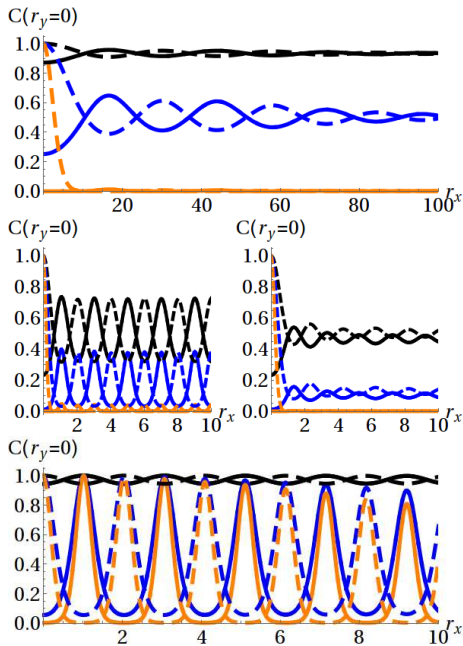}
\caption{Spatial dependence of concurrence $C$ when one TLS sits at $\{r_x,r_y\}$ distance from the other,
with initial state $|\Psi_-\rangle$ (dashed lines) or $|\Psi_+\rangle$ (continuous lines) possessing maximum entanglement $C=1$. The system-environment coupling is $\lambda=0.01\omega_0$. 
{\it Top:} isotropic case $\Omega=1.001\omega_0$ at times $\omega_0 t=10^3,10^4, 10^5$ (black, blue, orange). Note that solid and dashed coincide. 
{\it Middle:} `diagonal' case $\Omega=\sqrt{5/2}\omega_0$ at times $\omega_0 t=10^3,3\times10^3, 10^4$ (black, blue, orange).
{\it Bottom:} high-$k$ case $\Omega=1.999\omega_0$ at times $\omega_0 t=10^3,5\times10^4, 10^5$ (black, blue, orange).} 
\label{figSPINS}
\end{figure}

We analyze next how an initial entangled state of the two bosonic/TLS probes evolves
as a result of the interaction with an {\it infinite} lattice. We consider a two-mode squeezed vacuum \cite{2mode1,2mode2,2mode3}
for the oscillator case, and one of the two maximally entangled Bell states $|\Psi_\pm\rangle=(|01\rangle\pm|10\rangle)/\sqrt{2}$
for the two TLS case. To quantify entanglement for two TLS, we use the concurrence $C$ \cite{wootters} and for the bosonic case the logarithmic negativity \cite{vidalwerner}.
We start with the dynamics when the atom-lattice interaction is so weak that the master equation coefficients can be assumed to reach their constant asymptotic values, corresponding physically
to the regime where waves between atoms have enough time to become stationary (i.e. first and second Markov approximations). 
We show results for a couple of TLS for different times $\omega_0t$ (to be compared with the system-bath interaction time $1/\lambda^2=10^4/\omega_0$) and distances $\vec{r}$ 
in figure~\ref{figSPINS}, where we plot the surviving concurrence for initial states $|\Psi_\pm\rangle$. They follow the qualitative behavior of the cross-damping's spatial 
dependence, including the alternating behavior of $|\Psi_\pm\rangle$ as the sign of $\Gamma_c$ changes.  
\begin{figure}[h!]
\includegraphics[width=0.65\columnwidth]{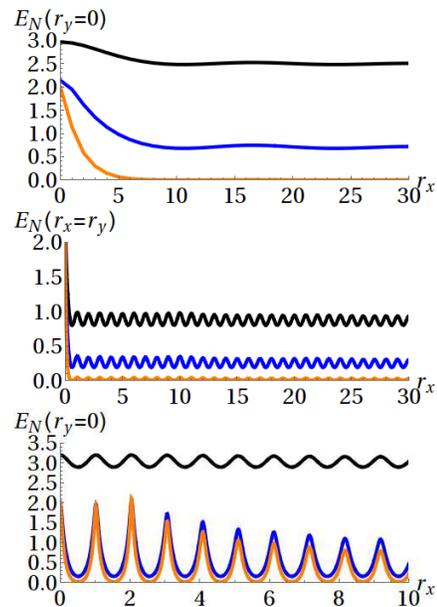}
\caption{Spatial dependence of logarithmic negativity $E_{\mathcal{N}}$ when one oscillator sits at $\{r_x,r_y\}$ distance from the other,
with an initial two-mode squeezed vacuum state with squeezing factor $r=2$; this state has $E_{\mathcal{N}}=4$. The parameters 
$\lambda$, $\omega_0 t$ and the cases are as in figure~\ref{figSPINS}.} 
\label{figOSCS1}
\end{figure}
\begin{figure}[h!]
\includegraphics[width=0.65\columnwidth]{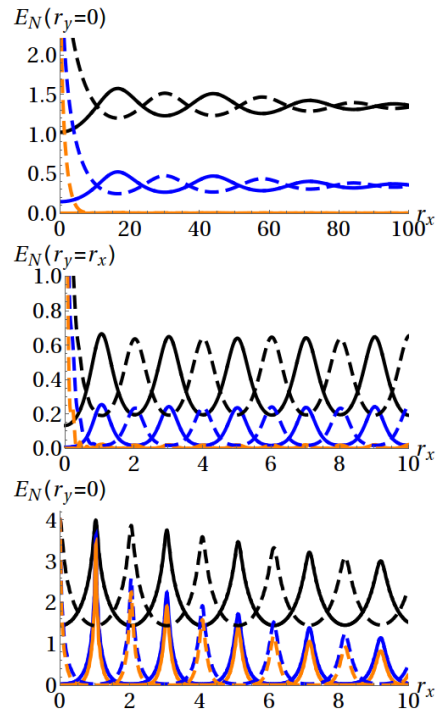}
\caption{Same as in Fig.~\ref{figOSCS1}, but initial state is asymmetrically squeezed (see text), with $(r_1,r_2)=(4,0)$ (continuous lines) and $(r_1,r_2)=(0,4)$ (dashed lines);
this state also has $E_{\mathcal{N}}=4$, but behaves more similarly to the atomic Bell states. The parameters 
$\lambda$, $\omega_0 t$ and the cases are as in figure~\ref{figSPINS}.} 
\label{figOSCS2}
\end{figure}
There is noiseless behavior (total subradiance) only if we are able 
to make $\vec{r}\ll 1/|\vec{k}_\Omega|$ (low {\it k}), for $|\Psi_-\rangle$, or with $\Omega$ at the high-{\it k} edge of the band placing the TLS at odd
($|\Psi_-\rangle$) or even ($|\Psi_+\rangle$) distances from each other. It must be noted though, that when the units are near the band edges they form atom-photon (or ion-phonon)
bound states which cannot be treated under Born-Markov conditions \cite{wang90,quang94,intJ}. The interesting `diagonal' case features $\Gamma_c(\vec{r})/\Gamma_0=1/2$ as maximum values, 
so it never produces noiseless dynamics.

The case of a bosonic two-mode state is, in comparison, insensitive to the distinction between CB and aCB, as can be seen in fig.~\ref{figOSCS1} (in some sense
each degree of freedom acts as each Bell state, but now both are present in the two-particle dynamics). This leads to similar although faster decay.
In order to see a behavior similar to atomic Bell states, we need to initialize the state as $|r\rangle_+\otimes|0\rangle_-$ (or $|0\rangle_+\otimes|r\rangle_-$), 
the equivalent of $|\psi_+\rangle$ (or $|\psi_-\rangle$, respectively), with the mode corresponding to operator $a_+$ squeezed and other in vacuum (or vice versa).
In comparison, the two-mode squeezed state is $|r\rangle_+\otimes|-r\rangle_-$. In this asymmetrically squeezed case, the dynamics is qualitatively similar to atomic Bell states, as can be seen
in Fig.~\ref{figOSCS2}.

\section{Finite size lattices} Finite size PC lattices typically need finite element calculations to compute the guided modes \cite{bookPHOT}, 
and are leaky at their boundaries. However we can use guided modes coupled to the atoms: these modes have evanescent tails at the outside of the crystal. 
Also, we could enforce Dirichlet boundary conditions for all allowed modes (i.e. also guided resonance modes which are within the light cone and thus can interact 
with the outer modes, allowing extraction of light \cite{JoanoGUIDED}) with Bragg mirrors at the boundaries \cite{Bragg}. Assuming that in practice this can be achieved
with enough precision, traveling waves are then replaced by standing waves
$$f_{\vec{n},\vec{k}}=\frac{2}{N+1}\sin\left(k_xn_x\right)\sin\left(k_yn_y\right)$$
with $k_{x,y}=\pi l_{x,y}/(N_{x,y}+1)$, $n_{x,y}\in[1,N_{x,y}]$ and the same for $l_{x,y}$. $N_{x,y}$ is given by the amount of dielectric function changes in each direction, typically associated with
holes made in a dielectric slab, or pillars above it \cite{bookPHOT}. For the finite size lattice we will take the tight-binding dispersion relation, which is as before 
$\omega_0-2J(\cos k_x +\cos k_y)$. A unitary transformation of the Hamiltonian removes the irrelevant offset $\omega_0$, which from now on we take as zero.
Now the damping coefficients become
$$\Gamma_0=2\lambda^2\sum_{k_x,k_y=1}^{N}\ 
\frac{\sin[t(\Omega-\omega_{\vec{k}})]}{\Omega-\omega_{\vec{k}}}|f_{\vec{n},\vec{k}}|^2$$
$$\Gamma_c=2\lambda^2\sum_{k_x,k_y=1}^{N}\ 
\frac{\sin[t(\Omega-\omega_{\vec{k}})]}{\Omega-\omega_{\vec{k}}}f_{\vec{n},\vec{k}}f_{\vec{n}',\vec{k}}$$
and accordingly the LS coherent contributions.
 \begin{figure}[h!]
\includegraphics[width=0.8\columnwidth]{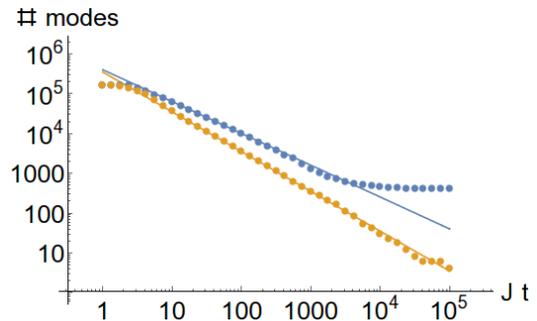}
\caption{Number of modes resonant (i.e. selected by the $sinc$ function, with $\Omega=0$(blue dots) and $\Omega=2J$(beige dots), as function of time with $N=400$. 
Fits with $t^{-1}$ (beige continuous line) and $t^{-0.8}$ (blue continuous line) are drawn for comparison.} 
\label{figmodes}
\end{figure}
\subsection{Number of resonant modes}
Once a discrete spectrum is assumed, consistent with the finite size of the lattice, one can understand the divergence in the decay rate as a scaling of 
the number of resonant modes as follows. Let us take for simplicity the case of periodic boundaries, so $f_{\vec{n},\vec{k}}=e^{-i\vec{k}\cdot\vec{n}}/2\pi$ 
and $|f_{\vec{n},\vec{k}}|^2=(2\pi)^{-2}$. The decay rate in this case is
\begin{equation}
\label{decayFINITE}
\Gamma_0(t)=\frac{2\lambda^2t}{(2\pi)^2}\sum_{\vec{k}}\rm{sinc}(t(\Omega-\omega_{\vec{k}})).
\end{equation}
For long times, this sum approximately counts how many modes in the crystal are resonant with $\Omega$, because for them the sinc function is 1, and 0 
elsewhere. In figure~\ref{figmodes} we can see that for frequencies $\Omega\neq0$, at the beginning all modes are resonant, but as time grows, the number of resonant modes 
quickly reaches a scaling $t^{-1}$, making the decay rate tend to a constant already for $J\cdot t\sim10$, consistent with figure~\ref{figDIBUJO}d.
This is a common situation, justifying the use of long-time master equations (Markov 2 approximation \cite{breuer} with constant coefficients.
However, for $\Omega=0$ we see that the number of resonant modes scales as $~t^{-0.8}$, and thus the decay rate scales as $t^{0.2}$. 
At long times it saturates at a value $4\times N/2$ and does not fall more, because the resonant manifold defined by the red square in fig.~\ref{figDIBUJO}b ($N/2$ modes in each 
of the four lines) perfectly fits the $sinc$ function in eq.~(\ref{decayFINITE}), whatever $t$. From this time on, the decay rate scales as $N\cdot t$, diverging for $N\to\infty$ as expected from a van Hove singularity.

\subsection{Multi-atom dark states} 
\begin{figure}[h!]
\includegraphics[width=\columnwidth]{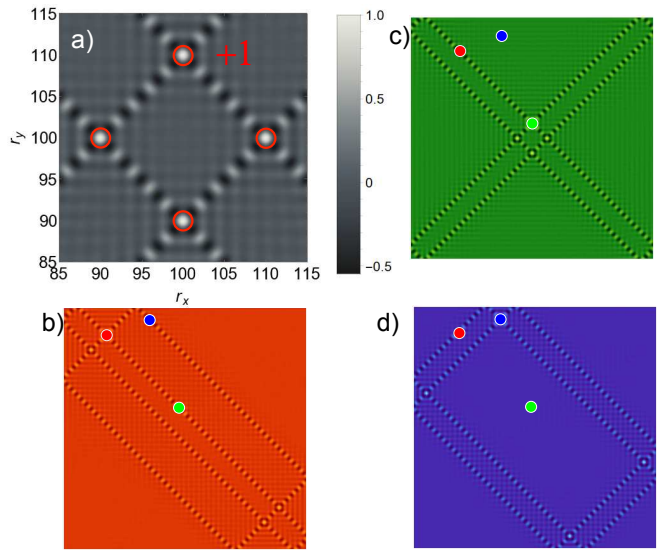}
\caption{`Diagonal' case $\Omega=\omega_0=0$. a) Cross-damping (in a $200\times 200$ lattice) e.g. between a particle at $(100,110)$ and another at $(r_x,r_y)$. The four points display
$\Gamma_c/\Gamma_0=+1$ (red). If we place one atom in each of these four points we get 3 completely subradiant and one superradiant states. Case with 3 atoms (in a $100\times 100$ lattice): 
probe 1 is green, probe 2 is red and probe 3 is blue; in b) we have plotted the cross-damping felt between red particle and another particle sitting at $(r_x,r_y)$, 
i.e. $\Gamma_c(red;r_x,r_y)/\Gamma_0$. In c) and d) the same quantity for green and blue particles, respectively.} 
\label{figFinite}
\end{figure}

Dark states (perfect subradiance) can arise among distant atoms in 1D but not in 2D isotropic environments.
In the following we show that the `diagonal' case with its linear radiation pattern can be used to 
 construct subradiant states, in analogy with the 1D case where $\Gamma_c^ {(1D)}/\Gamma_0=\cos(k_\Omega x)$ \cite{painter2015,Galve2016}, i.e. it makes $|\Psi_\pm\rangle$
 subradiant for specific distances between atoms due to coherent cancellation of waves. It is easy to see geometrically in an open 2D configuration that,
 because both atoms radiate with a cross pattern, only one of the lines (the one which joins their positions) can cancel. 
 As noted recently \cite{Tudela1,Tudela2}, no subradiant 2D configuration of two atoms exists in infinite planar settings. Geometrically again, we see that if we can 
 join the remaining two lines, total cancellation can occur \cite{galve2018}. Indeed this is the case for reflecting boundaries, as seen in Fig.~\ref{figFinite}a. 
 Only four positions for the two atoms will yield total cancellation (equivalently, $\Gamma_c/\Gamma_0= 1$, red points): placing two atoms on any of those points with a state 
 $|\Psi_-\rangle$ yields total subradiance, up to retardation effects. We stress that these dark states can be created at arbitrary distances, depending on the off-set from the center of the PC slab.

 The reflection of (photonic/phononic) waves can be
 used to build peculiar radiation patterns of multi-unit architectures. In Fig.~\ref{figFinite}b,c,d we show the cross-damping $\Gamma_c/\Gamma_0$ of one atom 
 with respect to other atom at position $(r_x,r_y)$: one unit (green) shares a partial subradiant condition ($\Gamma_c/\Gamma_0=1/2$) with another unit
 (red), the red unit shares partial subradiance with the blue unit ($\Gamma_c/\Gamma_0=1/2$), but the green and blue units radiate independently. 
 We have plotted the cross-talk of each of the atoms with the rest of the possible atomic positions in the crystal, to make it more understandable.
 In this particular case the decay rate matrix would be  $\Gamma_{i,j}=\{1,1/2,0\},\{1/2,1,1/2\}\{0,1/2,1\}\}$, with the slowest decay channel 
 corresponding to state $| \psi\rangle=(| 100\rangle-\sqrt{2}| 010\rangle+| 001\rangle)/\sqrt{2}$
 (with atom notation $|$green, red, blue$\rangle$) radiating ~5.8 and ~3.4 slower than the other two orthogonal states of three atoms.\\
 \indent Another interesting case is to put four atoms, e.g. each in the points highlighted in fig~\ref{figFinite}c (rhomboid configuration).
 This leads to $\Gamma_{i,j}=1$, $\forall i,j$ and thus to one superradiant state with decay $4\Gamma_0$, $|\psi_{\text{super}}\rangle=(|1000\rangle+|0100\rangle+|0010\rangle+|0001\rangle)/2$ 
 or simply $|\psi_{\text{super}}\rangle=\frac{1}{2}(1,1,1,1)$ (expressed in the basis $\{ |1000\rangle,|0100\rangle,|0010\rangle,|0001\rangle \}$, 
 with atom notation $|$up,bottom,left,right$\rangle$). Also to {\it completely subradiant} states of two (e.g. $(1,0,0,-1)/\sqrt{2}$), three (e.g. $(-1,0,2,1)/\sqrt{3}$)
 and four atoms (e.g. $(1,-3,1,1)/2\sqrt{3}$; see also \cite{galve2018}). Notice that one can keep several of the atoms in the ground 
 state and they will not affect the subradiance of the remaining excited atoms. In contrast, 
 in infinite lattices the decay rates we have the same superradiant state with lower decay rate $2\Gamma_0$, a totally 
 subradiant four atom state \cite{Tudela1,Tudela2} $|\psi_1\rangle=\frac{1}{2}(-1,-1,1,1)$, and the states of two or three atoms now radiate with $\Gamma_0$.
These examples show the potential of this platform to explore the exotic scenarios arising when all sites are filled with atoms in different architectures. 

\begin{figure}[h!]
\includegraphics[width=\columnwidth]{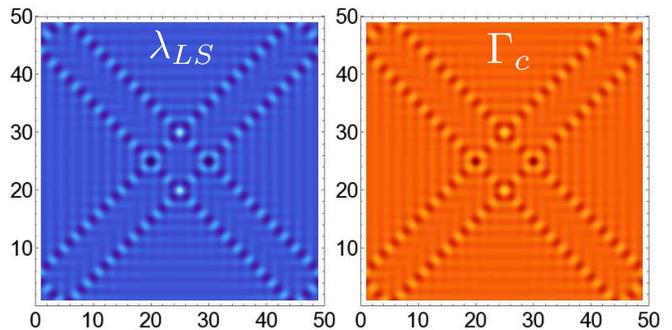}
\caption{Comparison of Lamb shift coupling $\lambda_{LS}$ (left), and cross-talk $\Gamma_c$ (right), with a tight-binding lattice $N_x=N_y=49$, at times $J\ t=10^5$. This corresponds to one of the 
cases studied in \cite{galve2018}. In absolute values $\lambda_{LS}\ll\Gamma_c$ though.} 
\label{figLamb}
\end{figure}
 
\subsection{Lamb-shift coupling}
One could worry about the stability of dark states because of the Lamb-shift coupling, which in principle could modify the phases of dark states turning them into bright.
For the most interesting `diagonal' case, and in contrast with the infinite lattice scenario, the values of $\lambda_{LS}$ are far smaller than the radiative coupling $\Gamma_c$.
In spite of both having the same spatial shape (see Fig.~\ref{figLamb}, now taking the tight-binding dispersion), we observe (not shown) that the ratio $\lambda_{LS}/\Gamma_c$ grows linearly with a rate $\sim 10^{-6}$, i.e.
it takes $J~ t\sim 10^5$ to reach $\lambda_{LS}/\Gamma_c\sim 0.1$. In addition, let us place four atoms in the four special points of maximum $\Gamma_c$ in a subradiant state. 
Because the matrix of Lamb-shift couplings $\lambda_{LS}(\vec{r}_i,\vec{r}_j)$ (with $\lambda_{LS}(\vec{r}_i,\vec{r}_i)=\Omega_{LS,i}$) has the same signs and structure as 
the cross-talk matrix $\Gamma_c(\vec{r}_i,\vec{r}_j)$, it is easy to show that
it does not perturb dark states \cite{galve2018}. This is because dark states are eigenstates of the cross-talk matrix with eigenvalue 0, and thus also of the Lamb-shift matrix.

\begin{figure}[h!]
\includegraphics[width=0.8\columnwidth]{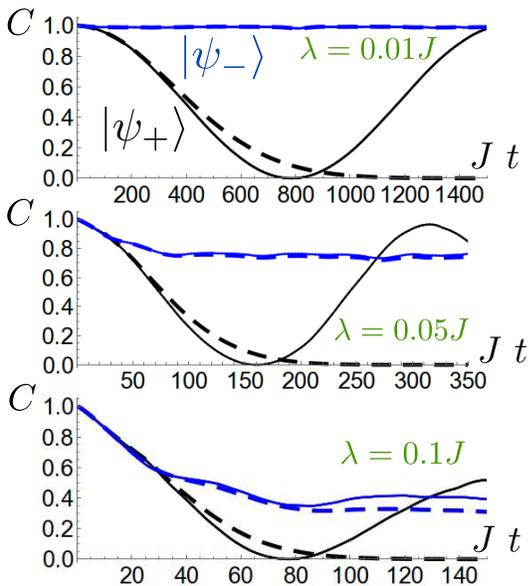}
\caption{Dynamics of concurrence of initial states $|\psi_{-(+)}\rangle$ in blue (black), with $\Omega$ at the middle of the band, as in Fig.~\ref{figLamb} (see main text). Continuous lines 
represent the exact dynamics, while dashed lines are the predictions of our time-dependent master equation. The system-lattice coupling is
$\lambda/J=0.01$ (top), $\lambda/J=0.05$ (middle) and $\lambda=0.1J$ (bottom). In the middle and bottom cases, a build-up time for the cross-talk is clearly observed.} 
\label{figCOMP}
\end{figure}

\subsection{Entanglement decay: perturbative vs. exact}
We finish by comparing the validity of our master equation with time-dependent coefficients and the exact dynamics of the model in the one-excitation sector \cite{galve2018,Tudela1}.
Take two atoms in the configuration of Fig.~\ref{figLamb}, at positions $\vec{r}_1=(25,20)$, $\vec{r}_2=(25,30)$, with initial Bell states $|\psi_\pm\rangle$. Since at those points $\Gamma_c/\Gamma_0=+1$,
we should, and do, observe no decay of the concurrence for $|\psi_-\rangle$ and fast decay for $|\psi_+\rangle$. While the shapes are not exactly the same, the master equation faithfully captures the time
it takes for the cross-talk to build-up. During this time, both Bell states decay because $\Gamma_c$ has not yet reached its final value. From that time on, only one of the Bell states decays further.
It is also interesting to see that, once the concurrence of $|\psi_+\rangle$ has fallen to zero, it can never go up. This is a clear effect of the Born approximation which takes the state of the lattice to 
be always the T=0 thermal state (vacuum). Because of this, no excitation can ever go back to the atoms.
We also observe that our master equation underestimates decay, as compared to the exact dynamics which gives oscillatory behavior (revivals). Finally, for system-lattice strong coupling $\lambda=0.1J$ we start 
observing more pronounced differences in the dynamics. This is a very interesting subject for future studies.\\
 
\section{Conclusions}
We have studied the interaction of two probes induced by a 2D structured periodic environment. 
The physics of this problem is generic enough that it can be applied to e.g. 2D Coulomb crystal of trapped 
ions \cite{thompson,Bollinger1,Bollinger2}, PCs with nearby trapped atoms \cite{painter2014,painter2015,chang2015,tudela2015}, 3D printed 
photonic circuits \cite{Sciarrino1,Sciarrino2}, trapped-ions near piezoelectric substrates \cite{giedke2015} or even circuit QED architectures \cite{cQED0,cQED1,cQED2,cQED3}, 
only requiring translational invariance of the lattice and weak system-lattice coupling. Coherent (Lamb Shift) and incoherent (radiative) 
couplings are studied both for finite and infinite lattices, analyzing many details and possible pitfalls. We show the usefulness of the master equation description, in terms of
radiative couplings between quantum units, by showing that it directly translates into the behavior of entanglement decay of two units. Further, for more than two units, we show
how to use the radiative coupling matrix to build dark states.

\section{Acknowledgements} We acknowledge fruitful discussions with A. Gonz\'alez-Tudela and J.I. Cirac.
This work has been supported by the EU 
through the H2020 Project QuProCS (Grant Agreement 641277), by MINECO/AEI/FEDER through projects NoMaQ FIS2014-60343-P, QuStruct FIS2015-66860-P
and EPheQuCS FIS2016-78010-P. FG acknowledges funding from `Vicerectorat d'Investigaci\'o  i Postgrau'  of the UIB.

\end{document}